\documentclass[12pt]{article}
\usepackage{graphicx}
\usepackage{cite}
\usepackage{amsmath}

\begin{document}


\title{Fake news: \\ ``No ban, No spread - with Sequestration"}

\author{ Serge Galam\thanks{serge.galam@sciencespo.fr} \\
CEVIPOF - Centre for Political Research, Sciences Po and CNRS,\\
1, Place Saint Thomas d'Aquin, Paris 75007, France}

\date{January 4, 2024}

\maketitle

\begin{abstract}

Fake news is today a major threat to free and democratic making of public opinion. To curb their spread, all efforts by institutions and policy makers rely mainly on imposing restriction, prohibition and fact checking sites, which end up to an effective limitation of freedom of speech. This policy of prohibition, supported by a wide consensus, has been recently broken by the controversial policy applied by Elon Musk to regulate the social media X, with a backlash accusing him of promoting hate speech. Here, notwithstanding these two policies,  I explore another avenue denoted ``No ban, No spread - with Sequestration", which amounts at the same time preserving full freedom of speech and neutralization of fake news impact. To investigate the feasibility of my proposal, I tackle the issue within the Galam model of opinion dynamics, a sociophysics stylized model, which has demonstrated an ability to predict a few unexpected political events including 2016 Trump election. In addition to the basic ingredients of the model, I explore for the first time the effect on the dynamics of opinion of a simultaneous activation of prejudice tie breaking and contrarian behavior. The results show that indeed most pieces of fake news do not propagate beyond small groups of people and thus pose no global threat. However, I have unveiled some peculiar sets of parameters for which fake news, even if initially shared by only a handful of agents,  spreads ``naturally" to invade a whole community with no resistance. Based on these findings, I am able to outline a path to neutralize such invasive fake news by blocking "naturally" its spread, effectively sequestering it in very small social networks of people. The scheme relies on reshaping the social geometry of the landscape in which fake news evolves. No prohibition is required with fake news left free to prosper but being sequestrated. Next challenging step will be designing measures to implement the model's findings into the real world of social media. 

\end{abstract}


\newpage
\section{Introduction}

Fake news has emerged as a significant and critical component of today information landscape. It plays a pivotal role in the shaping outcomes of critical debates surrounding social, political, and societal matters. Fake news exerts a substantial influence on the formation of public opinion via social media  \cite{social1, social2, social3}.

Frequently, fake news is crafted with malicious intent, disseminating false and deceptive information while employing emotional manipulation and exploring many prejudices deep rooted in our social and political representations.

The devastating impact of the fake news posted on October 17, 2023, by Hamas about the bombing of an hospital in Gaza, has illustrated the destructive power of fake news  \cite{fa1, fa2, fa3}. In particular, it has illustrated how social media have fostered the speed at which fake news can reach at once the entire world from common citizens to world leaders  \cite{fa4, fa5, fa6}. Indeed, the motivation of individual spreaders has been investigated  \cite{v1, v2}.

In addition, instances of foreign-originated fake news were observed during the 2016 US and 2022 French presidential elections, in which they interfered in favor of specific candidates. However, investigations have not established a quantifiable impact on the final electoral results, despite evidence pointing to a reinforcement of existing individual opinions  \cite{social4}.

Presently, fake news has evolved into a pervasive tool for distorting public discourse on crucial issues faced by modern societies. Consequently, its associated impact has become a pressing concern for democratic institutions. Policymakers are taking measures to implement various regulations aimed at curbing the spread of fake news. Considerable resources and effort are being dedicated to establishing fact-checking platforms to assess the reliability of information disseminated online. 

Nevertheless, fake news continues to thrive within social media platforms. Yet, it is important to note that not all fake news propagate beyond small networks of believers and thus, do not pose a significant threat. Nonetheless, here and there,  a few fake news spread over at high speed and propagate all over becoming worrying issues.

On this basis, with the phenomenon remaining basically unexplained despite current important theoretical and experimental efforts, to curb the destabilizing invasive fake news requires curbing all fake news including the many non-threatening ones. Indeed, the current trend among policy makers focuses on implementing overall control and regulation measures to restrict the use of social media  \cite{f1, f2, f3}, which in practical terms are likely to end in limitations of individual freedom of speech.

At odd with above trend, Elon Musk has launched an opposite controversial governance of his social media X, advocating for freedom of speech  \cite{f4, f5}, which in turn has produced a backslash from quite a number of people and institutions accusing him of promoting hate speech in the name of freedom of speech  \cite{f6}.

In this paper I address the issue of fake news by adopting another different view point I denote ``No ban, No spread - with Sequestration". The aim is to get fake news posts sequestrated within small networks of agents. Instead of attempting to curb pieces of fake news based on their respective contents, my starting hypothesis is to consider that it may not be the content of fake news per se that matters, but rather the interaction of that content with the social and psychological geometry in which fake news emerges and tries to propagate. While this geometry is given independently of the specific content of a fake news item, it plays a crucial role in facilitating or blocking its spread. 

Therefore, my focus is to address the reshaping of the associated social and psychological geometry to neutralize the structural features, which unlock the dissemination of some pieces of fake news, to make sure they stay confined and sequestrated  in small  networks of agents.

To test the soundness of my hypothesis, I consider a stylized social framework within sociophysics  \cite{brazil, frank, book, bikas}, that allows me to explore the features that control the propagation of fake news beyond its mere content. 

Sociophysics is a new active field of physics, which tackle social and political phenomena adopting a physicist-like approach  \cite{inter, nun1}. The challenge is not to substitute to social sciences but to create a new hard science by itself \cite{phys1, phys2, phys3, phys4, phys5, phys6, phys7, phys8}.

While sociophysics covers a large spectrum of social and political issues \cite{polari, cui, liu, banish, depo, decis, mau, nuno3, brics, success, rebel, shen, grabish, nuno4, roni1}, the study of the dynamics of opinion occupies a central place  \cite{mobile1, sen, zan, che, dispa, roni2 , andre1}. A good deal of papers consider binary variables  \cite{tot, mala, red, gm3, kas1, bol, bag, car,  kas2, flo, mar, igl, fas, gim, iac, bru, paw} and fewer three or more discrete opinions  \cite{mobilla, celia, andre2, malarz, voting3,  mobile2, entro}. 
 
Here, I extend the Galam model of opinion dynamics deployed in a multi-dimensional space of parameters, by exploring a novel combination of the heterogeneity of agents  \cite{gtak, chop, min, het, cont}. In particular, I investigate the effect of having some contrarians embedded among floaters with occurrence of tie breaking prejudice. 

The study revealed that the combination of these two ingredients is instrumental to shape the geometrical landscape, which either blocks fake news items or on the contrary unlocks others  fostering an overall spreading. I found that the activation of a small proportion of contrarians associated with a favorable tie breaking prejudice opens the path to a massive spreading of fake news, even when it started being believed by only a handful of agents.

The findings could serve as a foundation to design non-restrictive regulations, which could prevent ``naturally" any  invasion of social media by fake news. The related lever is the setting of a geometrical sequestration of fake news within small networks of agents. Then, neither prohibition nor restriction is required.

The rest of the paper is organized as follows: Section 2 reviews the Galam model of opinion dynamics while Section 3 introduces a new combination of contrarians and prejudice tie breaking in local group updates of opinion. The mechanisms locking or unlocking the spreading of fake news are unveiled in Section 4. Section 5 identifies new strategies to sequestrate fake news posts without prohibiting them. The Conclusion contains a summary of the main results.

\section{The Galam model of opinion dynamics}

\subsection{Floaters dynamics and local majority}

The bare Galam model considers two competing discrete opinions A and B within a homogeneous population of floaters. Floaters are agents holding an opinion. They advocate to promote it. However, floaters listen to opposite arguments in favor of the competing opinion and thus are susceptible to get convinced to shift opinion \cite{chop, min, het}. 

Given initial proportions $p_0$ and $(1-p_0)$ in favor of A and B, a dynamic is implemented by iterating a three-step procedure to update individual opinions. First, agents are distributed randomly in small groups of size $r$. Then, a local majority rule is applied separately to each group where all agents adopt the majority opinion. Third, agents are reshuffled. The update modifies the proportions $p_0$ and $(1-p_0)$ to new ones $p_1$ and $(1-p_1)$. 

The scheme is then repeated $n$ times with $p_0 \rightarrow p_1  \rightarrow p_2  \rightarrow ...\rightarrow p_n$. The associated update equation is given by,
\begin{equation}
p_{r,1}=   \left [ \sum_{l=\bar r+1}^{r}   {r \choose l} p_{0}^l  (1-p_0)^{r-l} +\frac{1}{2}  \delta [\bar r-\frac{r}{2}]  {r \choose r/2}  p_{0}^{\frac{r}{2}}  (1-p_{0})^{\frac{r}{2}} \right ]  \ ,
\label{r} 
\end{equation}
where $\bar r \equiv I [\frac{r}{2}]$ with $I [...]$ meaning integer part of $[...]$ and $\delta [x]$ is the Kronecker function. Last term of Eq. (\ref{r}) means that for even value of $r$ at a tie, agents do not shift opinion. 

The full landscape of the dynamics is obtained solving the fixed point equation $p_{r,1}=p_{0}$. The equation yields one tipping point $p_t =\frac{1}{2}$ and two attractors $p_A=1$ and $p_B=0$,  which are all independent of the value of $r$. The associated dynamics is thus perfectly balanced with the opinion, which has gathered the majority of individual initial opinions, becoming larger and larger to reach eventually unanimity provided the number of updates $n$ is sufficient. 

Majority agents have convinced individually via local discussing groups, agents holding initially the minority opinion to adopt the initial majority opinion. The dynamics is democratic with $p_0 < p_1  < p_2  < ...< p_n$ when $p_0>\frac{1}{2}$ and $p_0 > p_1  > p_2  > ...> p_n$ when $p_0<\frac{1}{2}$.

\subsection{Floaters with tie breaking prejudice}

However, above ideal picture of a democratic opinion dynamics breaks down when a tie breaking is included for even size groups. In this case, at a tie the group get trapped into a collective doubt with both opinions being equally acceptable. Since rationality cannot help to decide, everyone selects one of the two opinions at random like with tossing a coin. Individual choices are made by chance. 

But contrary to above rational of random choices, the Galam model hypothesizes that indeed the ``coin" is biased. For each agent, the state of doubting puts unconsciously some prejudice in control of the choice. The decision making bias is monitored by the prejudice, which has been activated by the issue at stake. The decision is not made in the name of prejudice, but in the name of chance. To account for a distribution of different prejudices among agents, at a tie, opinion A is selected with probability $k$ and opinion B with probability $(1-k)$.

Accordingly, for $r=4$ Eq. (\ref{r}) reduces to, 
\begin{equation}
p_{4,k}= p_0^4+4p_0^3(1-p_0)+6k p_0^2(1-p_0)^2 ,
\label{4k} 
\end{equation}
which still has the two attractors  $p_A=1$ and $p_B=0$. But now, the attractors are separated by a tipping point located at,
\begin{equation}
p_{t,k}=\frac{(6k-1)-\sqrt{13-36k+36k^2}} {6(2k-1)} ,
\label{ptk} 
\end{equation}
instead of $p_t=\frac{1}{2}$. For $k=0, \frac{1}{2}, 1$, Eq. (\ref{ptk})  yields respectively $p_{t,0}=\frac{1+\sqrt{13}} {6} \approx 0.77$,  $p_{t,\frac{1}{2}}=\frac{1}{2}$ and $p_{t,1} = \frac{5-\sqrt{13}} {6} \approx 0.23$. Accordingly, $\frac{1}{2} \leq p_{t,k}\leq  0.77$ when $0\leq k \leq \frac{1}{2}$  and $0.23\  \leq p_{t,k}\leq \frac{1}{2}$ when $\frac{1}{2}\leq  k \leq 1$.\footnote{Note some misprints in Eq. (3) and following paragraph in S. Galam, Entropy 25(4), 622 (2023).}

With $p_t \approx 0.23$ instead of $0.50$, the case $k=1$ illustrates the phenomenon of minority spreading. Opinion A being favored by the group prejudice, it needs to gather at minimum an initial minority support of only 0.23\% to convince the initial majority of agents, who are sharing opinion B ,to adopt instead opinion A via local and open mind discussions. 

The previous democratic character of the opinion dynamics has been broken naturally and unconsciously without notice in favor of the choice in tune with the prejudices of the group. No coercion has been used. However, the initial support of A must be larger than $0.23$. Otherwise, the minority opinion does lose support.

\subsection{Floaters and contrarians}
\label{flo-con}

Like floaters, contrarians are agents having an opinion, arguing for it and  listening to opposite arguments. However, unlike floaters, instead of following the local majority in a discussing group (Fake news is true, fake news is false), they automatically adopt the opposite opinion (fake news is false, fake news is true) whatever the majority is \cite{cont}. They do it also in case of initial unanimity.

On this basis, while local majority rule favors the strengthening of an initial majority between two competing opinions, contrarians on the other hand favor a minority stand, which in turn reduces the gap between the respective proportions of the two competing opinions. 

As expected contrarians prevent the disappearance of the minority opinion even when may successive updates have been implemented. For instance, for discussing groups of size $4$, update Eq. (\ref{4k}) becomes,
\begin{equation}
p_{1,x}= (1-2x) \left [p_0^4+4p_0^3(1-p_0)+3 p_0^2(1-p_0)^2\right] +x ,
\label{4x} 
\end{equation}
where $k=\frac{1}{2}$ (no tie breaking), $x$ is the proportion of contrarians and $(1-x)$ the proportion of floaters.

The associated fixed point equation $p_{1,x}=p_0$ yields $p_{t,x}=\frac{1}{2}$ and,
\begin{equation}
p_{{A, x};{B,  x}}=\frac{1 - 2 x \pm \sqrt{1 -8 x+12 x^2}}{2 (1 - 2 x)} .
\label{pABx} 
\end{equation}
provided $0\leq x  \leq x_c$ with $x_c=\frac{1}{6} \approx 0.167$. 

While the tie prejudice effect preserves the floater attractors $p_A=1$ and $p_B=0$ shifting the tipping point away from $p_t=\frac{1}{2}$, contrarians on the opposite, preserve the tipping point $p_t=\frac{1}{2}$ but shift both attractors, which in turn stabilize a coexistence of a large majority and a small minority with $p_{A, x}<1$ and $p_{B, x}>0$.

It is worth noticing that at $x  = x_c=\frac{1}{6}$, $p_A=p_B=\frac{1}{2}$. Accordingly, for $x  > x_c$, the dynamics is driven by a single attractor located at $\frac{1}{2}$. Any initial condition ends up at a perfectly balanced support for A and B. Here, I assume that $x  \leq \frac{1}{2}$, which is sound given the definition of a contrarian.

\subsection{From minority opinion to fake news}

Galam model deals with competing opinions A and B given some initial proportions $p_0$ and $(1-p_0)$ of respective supports, making by definition one minority and the other one majority. To apply the model to fake news is done naturally noticing a restriction of the proportions. Denoting A fa piece of fake news implies by nature of the phenomenon to have A being ultra minority with $p_0<< \frac{1}{2}$. The competing opinion B then denotes the dismissing of A for being false and always starts with an overwhelming support. 

In addition it is worth to stress that each fake news post activates specific prejudices. Given the content, within a social and political context, a fake news item produces contrarians in different proportions.

\section{Contrarians with tie prejudice breaking} 

In above Section I reviewed some main results obtained from the Galam model of opinion dynamics. I now investigate for the first time the impact of having contrarians in a community of floaters with active tie prejudice breaking. I restrict the study to discussing groups of size 4. Combining Eqs.  (\ref{4k}) and (\ref{4x}) leads to the update equation,
\begin{equation}
p_{4,k,x}= (1-2x) \left [p_0^4+4p_0^3(1-p_0)+6k p_0^2(1-p_0)^2 \right]  + x .
\label{4kc} 
\end{equation}

Despite Eq.  (\ref{4kc}) apparent simplicity, the related fixed point equation $p_{4,k,x}= p_0$ cannot be solved analytically. A numerical treatment is required to determine the fixed points $p_{A, k, x}, p_{B, k, x}, p_{t, k, x}$ with their domains of existence within the full parameter space $0\leq k \leq 1$ and $0\leq x \leq 1$.

\begin{figure}[t]
\includegraphics[width=0.9\textwidth]{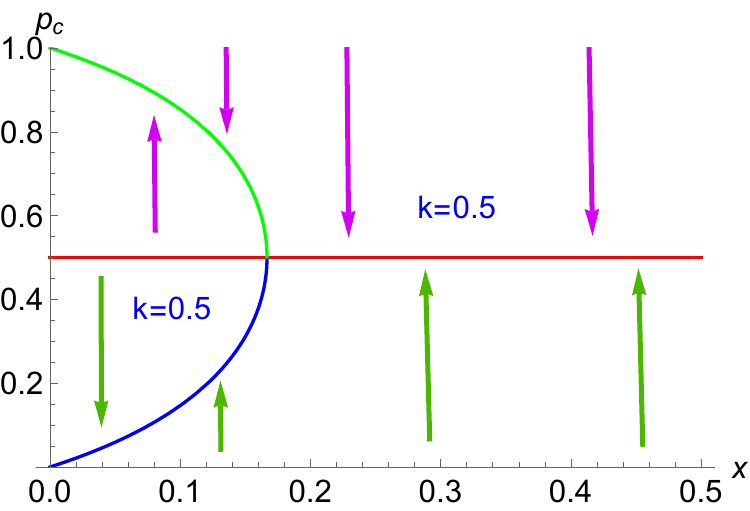}
\caption{Evolution of attractors and tipping points as a function of the proportion $x$ of contrarians for $k=\frac{1}{2}$. The lower curve represents the attractor $p_{B, 0.5, x}$ (in blue), the upper curve the attractor $p_{A, 0.5, x}$ (in green and red) and the middle curve the tipping point $p_{t, 0.5, x}=0.5$ (in red). At $x_c=0.055$, $p_{A, 0.5, x_c}=_{B, 0.5, x_c}=_{t, 0.5, x_c}=0.5$. For $x>x_c$ the unique attractor is located at precisely $\frac{1}{2}$.}
\label{p5}
\end{figure}

Before starting the investigation, it is worth to remind the effect on the opinion dynamics of having contrarians among a community of floaters without tie breaking prejudice ($k=\frac{1}{2}$) as shown in Fig.(\ref{p5}) and discussed in Subsection (\ref{flo-con}). With contrarians opposing the local majorities, as expected they prevent reaching unanimity securing always a resilient proportion of agents sharing the minority opinion. The more contrarians, the larger the stable minority with both attractors moving towards the tipping point, which keeps located at $50\%$. 

Nevertheless, despite being smooth, the shift of attractors bring them at the tipping point for the small proportion of contrarians $x_c=\frac{}{}\approx 0.167$. Then, for $x>x_c$, the dynamics is turned upside down becoming a single $50\%$ attractor dynamics. Beyond $x_c$, contrarians erase totally any initial difference in the proportions of support for A and B. The whole related dynamics is and stays symmetric ensuring a democratic balance.

However, as soon as $k\neq 0$, the symmetry between A and B is broken. To identify the consequences associated with the activation of contrarians I choose arbitrarily to start from $k=1$ to study the range $0\leq k \leq 1$.

\subsection{Fake news in tune with active prejudices}

\begin{figure}[t]
\includegraphics[width=.90\textwidth]{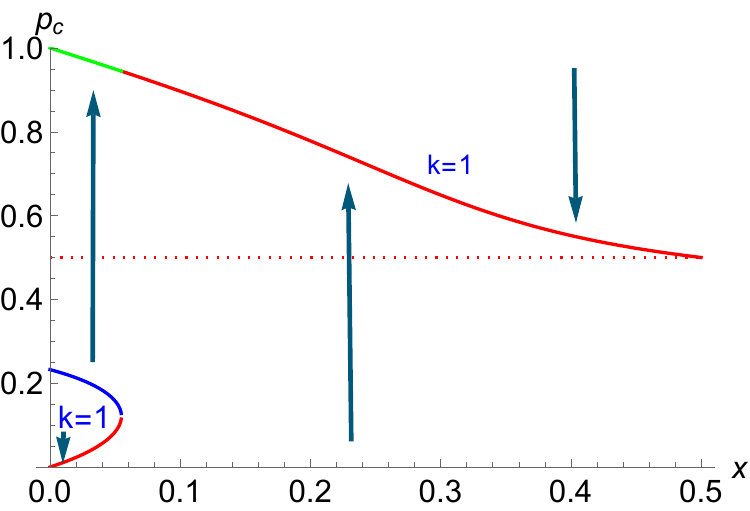}
\caption{Evolution of attractors and tipping points as a function of the proportion $x$ of contrarians for $k=1$. The lower curve represents the attractor $p_{B, 1, x}$ (in red), the upper curve the attractor $p_{A, 1, x}$ (in green and red) and the middle curve the tipping point $p_{t, 1, x}$ (in blue). At $x_c=0.055$, $p_{A, 1, x_c}=0.944$ and $p_{A, 1, 0.20} = 0.778$ and $p_{A, 1, 0.30} = 0.649$.}
\label{p1}
\end{figure}

Fig. (\ref{p1}) exhibits the dynamics landscape of opinion as a function of $x$ for $k=1$. At the corner $(k=1, x=0)$,  $p_{A, 1, 0}=1,  p_{B, 1, 0}=0$ and  $p_{t, 1, 0}\approx 0.23$. From there, increasing a bit the proportion $x$ of contrarians shifts $p_{A, 1, 0}$  and $p_{B, 1, 0}$ to respective lower and higher values $p_{A, 1, x}$  and $p_{B, 1, x}$ as expected intuitively and shown in the Figure.

However, the value of the tipping point $p_{t, 1, x}$ is seen to decrease, which is counter-intuitive. Indeed, being in a region where $p$ is less than $0.23$, a lower value for the tipping point means that A will keep increasing even when above fifty percent. That is counter-intuitive since contrarians are expected to decrease any majority as seen with the two attractors $p_{A, 1, 0}=1$ and  $p_{B, 1, 0}=0$ being shifted to respective lower and hight values.

Moreover, contrary to the symmetric case where the three fixed points merge to result in one unique attractor located at $0.50$, here only the two fixed points $p_{B, 1, x}$ and $p_{t, 1, x}$ merge and disappear at a small value $x_c=0.055$, leaving $p_{A, 1, x}$ as the unique attractor driving the dynamics with $p_{A, 1, x_c}=0.944$. It is noticeable that when $x>x_c=0.055$, $p_{A, 1, x}$ is always located above 50\%.

In addition to being counter-intuitive, above results show a surprising and worrying reality about the spreading of fake news. Given a fake news item totally in tune $(k=1)$ with the leading prejudice of a community, as soon as the proportion of active contrarians is larger than a few percents $(x>x_c=0.055)$, a handful of agents sharing initially the fake news is sufficient to have it spread inexorably and invade large part of this community as seen with $p_{A, 1, x_c}=0.944$. 

Although increasing the proportion of contrarians decreases the value  $p_{A, 1, x}$ towards fifty-fifty as expected, the fake news can still reach more than the majority of the community as seen with  $p_{A, 1, 0.20} = 0.778$ and $p_{A, 1, 0.30} = 0.649$ in Fig. (\ref{p1}).

\begin{figure}[t]
\includegraphics[width=.90\textwidth]{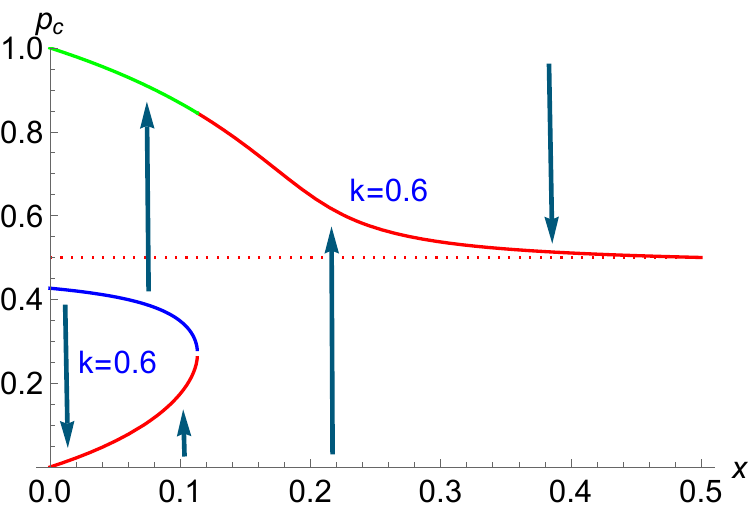}
\caption{Evolution of attractors and tipping point as a function of the proportion $x$ of contrarians for $k=0.60$. The lower curve represents the attractor $p_{B, 0.60, x}$ (in red), the upper curve the attractor $p_{A, 0.60, x}$ (in green and red) and the middle curve the tipping point $p_{t, 0.60, x}$ (in blue). At  $x_c=0.114$, $p_{A, 0.60, x_c}=0.843$ and $p_{A, 0.60, 0.20} = 0.648$ and $p_{A, 0.60, 0.30} = 0.537$.}
\label{p2}
\end{figure}

In case the prejudices are heterogeneous with respect to the fake news post, above phenomenon persists requiring only a little more active contrarians. But the fake news post still requires only a handful of initial support agents to spread over and become majority. For instance with $k=0.60$, $x_c=0.114$ and $p_{A, 0.60, x_c}=0.843$ as seen in Fig. (\ref{p2}). 

For $x=0.20$,  $p_{A, 0.60, 0.20} = 0.648$, which indicates that a large minority $1-p_{A, 0.60, 0.20} = 0.352$ remain dismissing the piece of fake news. For $x=0.30$, $p_{A, 0.60, 0.30} = 0.537$ with the opposed minority reaching $1-p_{A, 0.60, 0.30} = 0.463$.

\begin{figure}
\includegraphics[width=.50\textwidth]{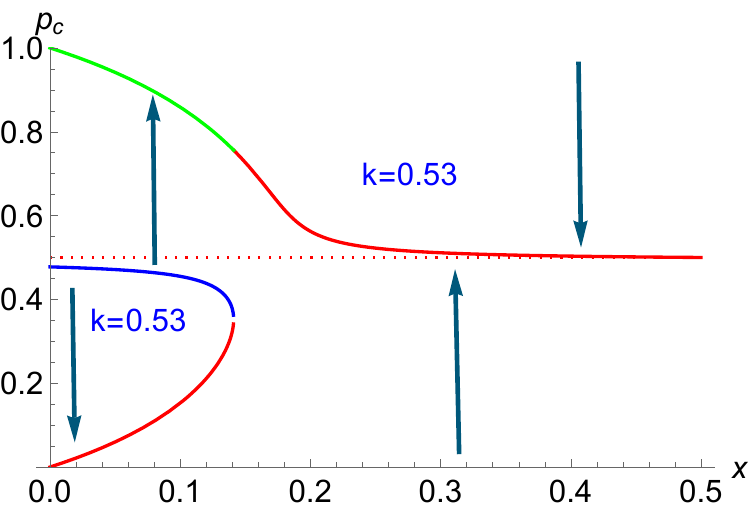}\quad
\includegraphics[width=.50\textwidth]{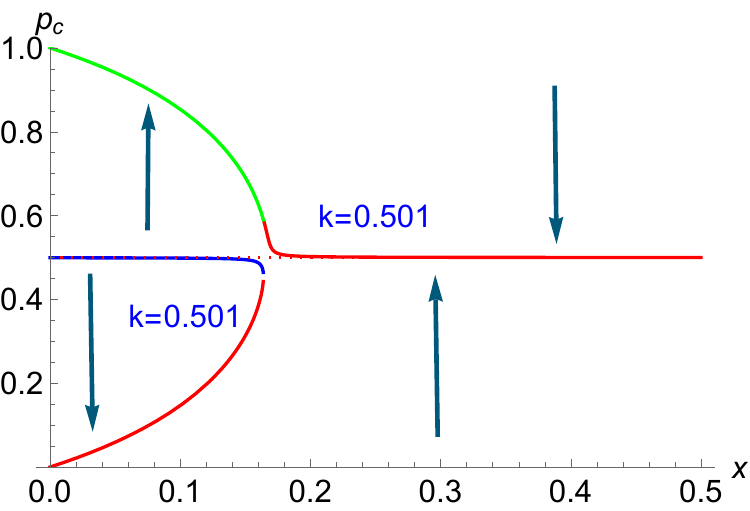}
\caption{Evolution of attractors and tipping points as a function of the proportion $x$ of contrarians for $k=0.53$ (left) and $k=0.501$ (right). The lower curves represent the attractor $p_{B, k, x}$ (in red), the upper curves the attractor $p_{A, k, x}$ (in green and red) and the middle curves the tipping point $p_{t, k, x}$ (in blue). On the left part  $x_c=0.142$, $p_{A, 0.53, x_c}=0.753$ and $p_{A, 0.53, 0.20} = 0.562$ and $p_{A, 0.53, 0.30} = 0.511$. On the right part  $x_c=0.164$, $p_{A, 0.501, x_c}=0.589$ and $p_{A, 0.501, 0.20} = 0.502$ and $p_{A, 0.501, 0.30} = 0.500$.}
\label{p34}
\end{figure}

Fig. (\ref{p34}) exhibits both cases $k=0.53$ and $k=0.501$, which show how the symmetrical case $k=0.50$ is recovered (see Fig. (\ref{p1})). The left part has $x_c=0.142$, $p_{A, 0.53, x_c}=0.753$, $p_{A, 0.53, 0.20} = 0.562$ and $p_{A, 0.53, 0.30} = 0.511$. The right part has $x_c=0.164$, $p_{A, 0.501, x_c}=0.589$, $p_{A, 0.501, 0.20} = 0.502$ and $p_{A, 0.501, 0.30} = 0.500$.

\subsection{Fake news at odds with active prejudices}

\begin{figure}[t]
\includegraphics[width=.90\textwidth]{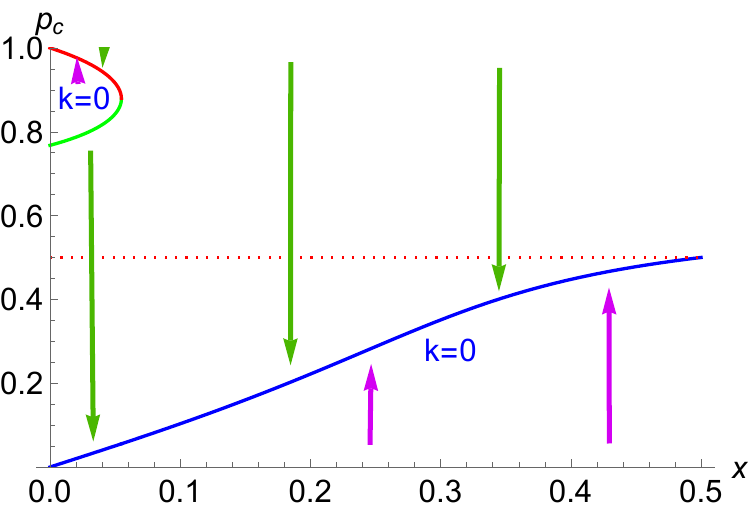}
\caption{Evolution of attractors and tipping points as a function of the proportion $x$ of contrarians for $k=0$. The lower curve represents the attractor $p_{B, 0, x}$ (in blue), the upper curve the attractor $p_{A, 0, x}$ (in red) and the middle curve the tipping point $p_{t, 0, x}$ (in green). At $x_c=0.055$, $p_{B, 0, x_c}=0.056$ and $p_{B, 0, 0.20} = 0.222$ and $p_{B, 0, 0.30} = 0.351$.}
\label{q1}
\end{figure}

I now look at the reverse situation with a fake news item at odds with the leading prejudices of the community. The extreme case $k=0$ dynamics landscape is shown in Fig.  (\ref{q1}). Comparing  Figs.  (\ref{q1}) and (\ref{p1}) shows that the situation is anti-symmetrical from $k=1$ where to be in tune with the leading prejudices fosters drastically the spreading of fake news even when initially shared by only a handful of agents.

When the fake news item is at odds with the active prejudices, even if a large majority of agents has initially believed it is true, the repeated discussions between agents in small groups reduce drastically their proportion. Fig. (\ref{q1}) exhibits the dynamics landscape as a function of $x$ for $k=0$. At the corner $(k=0, x=0)$,  $p_{A, 0, 0}=1,  p_{B, 0, 0}=0$ and  $p_{t, 0, 0}\approx 0.77$. From there, increasing a bit the proportion of contrarians shifts $p_{A, 0, 0}$  and $p_{B, 0, 0}$ to respective lower and higher values as expected intuitively. 

But here, contrary to $p_{t, 1, x}$ the value of the tipping point $p_{t, 0, x}$ increases with $x$, which is expected since a higher value for the tipping point makes more difficult for A majority to hold its status with contrarians reducing the gap between the majority and the minority. However, as above, even when A turns minority, it keeps losing support.

Now, the two fixed points $p_{A, 0, x}$ and $p_{t, 0, x}$  instead of $p_{B, 1, x}$ and $p_{t, 1, x}$, merge and disappear but still at the same small value $x_c=0.055$, leaving $p_{B, 0, x}$ to be the unique attractor driving the dynamics with $p_{B, 0, x_c}=0.056$. 

While above results were worrying and surprising, here the results are reassuring with respect to the spontaneous curbing of fake news diffusion. Given a fake news item totally at odds $(k=0)$ with the leading prejudice of a community, as soon as the proportion of active contrarians is larger than a few percents $(x>x_c=0.055)$, even a huge majority of agents sharing initially the fake news item will eventually shift opinion making their proportion to shrink inexorably down to low values of believers as seen with $p_{B, 0, x_c}=0.056$.

Moreover, increasing the proportion of contrarians increases the value  $p_{B, 0, x_c}$ towards fifty-fifty as expected but keeps it lower than $0.50$. Indeed when the fake news supporters are majority, the dynamics turn them down to a minority as seen with  $p_{B, 0, 0.20} = 0.222$ and $p_{B, 0, 0.30} = 0.351$ in Fig. (\ref{q1}).

\begin{figure}[t]
\includegraphics[width=.90\textwidth]{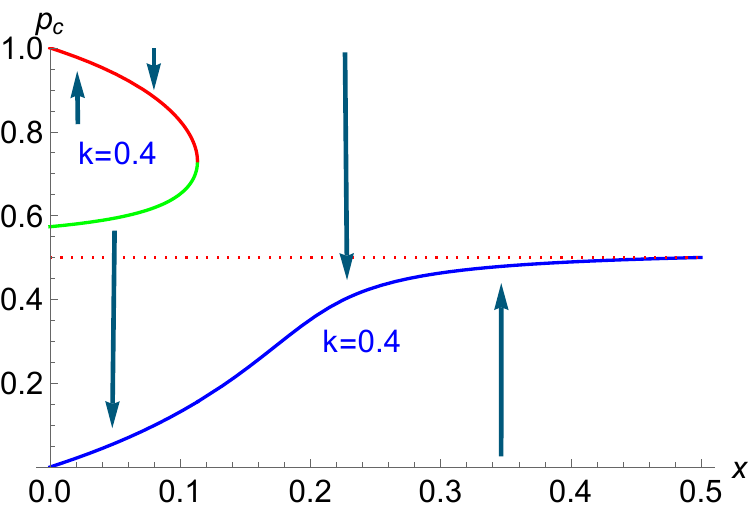}
\caption{Evolution of attractors and tipping points as a function of the proportion $x$ of contrarians for $k=0.40$. The lower curve represents the attractor $p_{B, 0.40, x}$ (in blue), the upper curve the attractor $p_{A, 0.40, x}$ (in red) and the middle curve the tipping point $p_{t, 0.40, x}$ (in green). At  $x_c=0.114$, $p_{B, 0.40, x_c}=0.157$ and $p_{B, 0.40, 0.20} = 0.352$ and $p_{B, 0.40, 0.30} = 0.463$.}
\label{q2}
\end{figure}

In case the prejudices are heterogeneous with respect to the fake news item, above phenomenon persists requiring only a little more active contrarians. The fake news item still ends up to a minority. For instance with $k=0.40$, $x_c=0.114$ and $p_{B, 0.40, x_c}=0.157$ as seen in Fig. (\ref{q2}). Nevertheless, Fig. (\ref{q2}) shows that a large minority remain believing the fake news item with $p_{B, 0.40, 0.20} = 0.352$ and $p_{B, 0.40, 0.30} = 0.463$ for respectively $x=0.20$ and $x=0.30$.

\begin{figure}
\includegraphics[width=.50\textwidth]{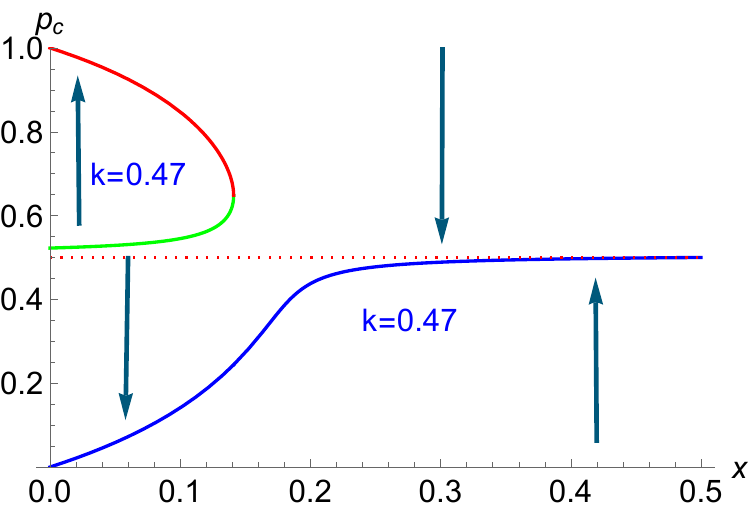}\quad
\includegraphics[width=.50\textwidth]{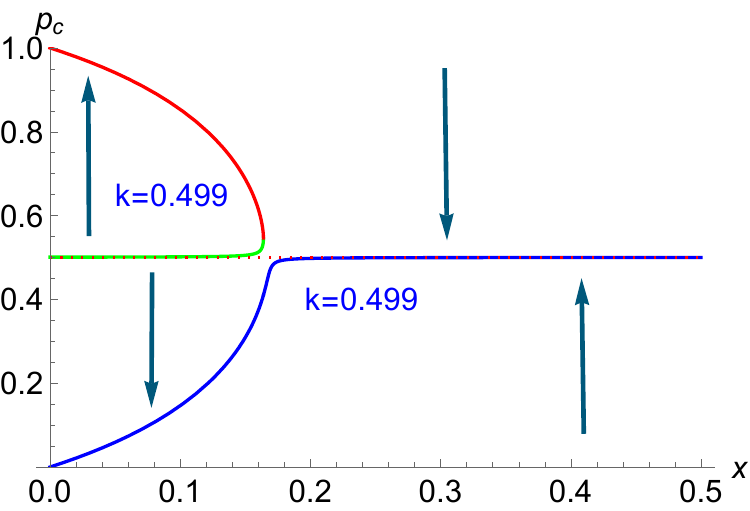}
\caption{Evolution of attractors and tipping points as a function of the proportion $x$ of contrarians for $k=0.47$ (left) and $k=0.499$ (right). The lower curves represent the attractor $p_{B, k, x}$ (in blue), the upper curves the attractor $p_{A, k, x}$ (in red) and the middle curves the tipping point $p_{t, k, x}$ (in green). On the left part  $x_c=0.142$, $p_{B, 0.47, x_c}=0.247$ and $p_{B, 0.47, 0.20} = 0.438$ and $p_{B, 0.47, 0.30} = 0.489$. On the right part  $x_c=0.164$, $p_{B, 0.499, x_c}=0.411$ and $p_{B, 0.499, 0.20} = 0.498$ and $p_{B, 0.499, 0.30} = 0.500$.}
\label{q34}
\end{figure}

Fig. (\ref{q34}) exhibits both cases $k=0.47$ and $k=0.499$ showing how the symmetrical case $k=0.50$ is recovered from lower value of $k$ (see Fig. (\ref{p1})). The left part has $x_c=0.142$ and $p_{B, 0.47, x_c}=0.247$. For respectively $x=0.20$ and $x=0.30$, $p_{B, 0.47, 0.20} = 0.438$ and $p_{B, 0.47, 0.30} = 0.489$. The right part has $x_c=0.164$ and $p_{B, 0.499, x_c}=0.411$ with $p_{B, 0.499, 0.20} = 0.498$ and $p_{B, 0.499, 0.30} = 0.500$.

\section{Unlocking or locking fake news}

Above analyses have highlighted the critical impact of a few percent of contrarians on the dynamics of spreading of a piece of  fake news, once tie breaking by prejudice is activated. In addition, the direction of the contrarian impact is set by the overlap between the content of fake news and the leading prejudices prevailing in the social community.  

\begin{figure}
\centering
\includegraphics[width=1\textwidth]{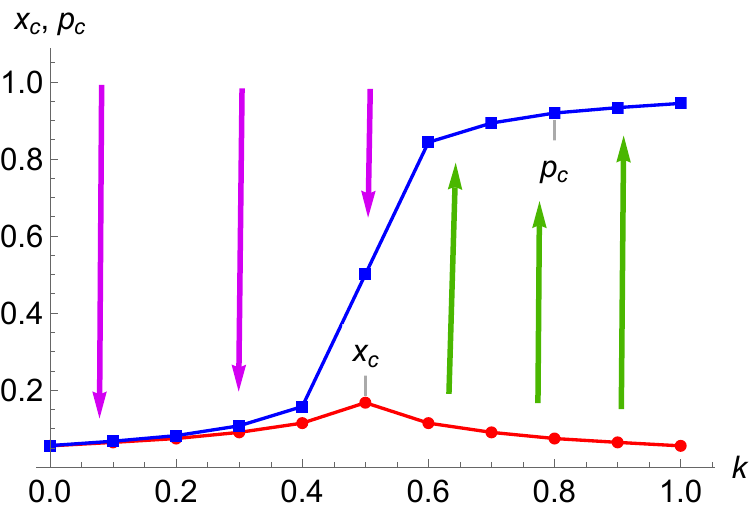}\quad
\caption{The values of $x_c$ as a function of $k$. The label $p_c$ denotes the values $p_{B, k, x_c}$ for $0 \leq k \leq 0.5$ and $p_{A, k, x_c}$ for $0.5 \leq k \leq 1$.}
\label{px}
\end{figure}

It is useful to reemphasize that the selection of the prejudice being activated  by the fake news is done unconsciously when a discussing group gets trapped in a local doubt determining the fake news validity. Either, the fake news item benefits from the leading activated prejudices with $k>\frac{1}{2}$ or it is impeded with $k<\frac{1}{2}$. The respective effects on the dynamics of spreading are drastically different. But in both case, only a few percent of contrarians are required to implement the drastic bias of the dynamics as shown in Fig. (\ref{px}) and Table (\ref{tt}).

\begin{table}
\begin{center}
\begin{tabular}{|c|c|c|c|c|c|c|c|c|c|c|} 
 \hline
 k & 0, 1 & 0.1, 0.9 & 0.2, 0.8 & 0.3, 0.7 & 0.4, 0.6 & 0.5 \\ 
 \hline
$x_x$ & 0.055& 0.064& 0.074& 0.09& 0.114& 0.167 \\
 \hline
$p_{B, k\leq 0.5, x_c}$ & 0.056& 0.067& 0.0815& 0.107& 0.157& 0.5  \\
\hline
$p_{A, k\geq 0.5, x_c}$ & 0.944& 0.933& 0.919& 0.893& 0.843& 0.5  \\
\hline
\end{tabular}
\end{center}
\caption{The values of $x_c$ as a function of $k$ with the values $p_{B, k, x_c}$ for $0 \leq k \leq 0.5$ and $p_{A, k, x_c}$ for $0.5 \leq k \leq 1$.}
\label{tt}
\end{table}

Moreover, the values of $x_c$ are identical for $k$  and $(1-k)$ in the range $0\leq k \leq \frac{1}{2}$. The associated values of the unique attractor at $x_c$ satisfy   $p_{B, k\leq 0.5, x_c} + p_{A, k\geq 0.5, x_c} =1$ as seen in Table (\ref{tt}). In addition it is worth noticing that the values of the unique attractors at $x_c$ stay either very low ($0 \leq k <0.4$) or very high ($0.6 < k \leq 1$) beside in the range $0.4 <k <0.6$ where the values respectively falls towards $0.5$. Three very different regimes are thus obtained as a function of $k$:

\begin{description}

\item[Regime 1 with $0 \leq k <0.4$:]  when the activated prejudices are mainly detrimental to fake news, the unique attractor is $p_{B,k,x}$ is always much lower than $\frac{1}{2}$ as seen in Table (\ref{tt}). It means that even if a fake news item is first believed to be true by an overwhelming majority of the agents, the subsequent informal discussions among small groups of agents will eventually turn most of them to reject the fakes news item as being false. 

\item[Regime 2 with $0.4 <k <0.6$:] when the activated prejudices are almost equally distributed with respect to fake news, whatever initial conditions, the fake news post ends up being shared by almost half of the agents. It is less than half when $0.4 <k <0.5$ and more than half for $0.5 <k <0.6$. In both cases, a substantial part of the community believes the fake news is true against another substantial part which believes it is false. The society is polarized with respect to the validity of the piece of fake news.

\item[Regime 3 with $0.6 < k \leq 1$:]  when the activated prejudices are mainly at the benefit of the fake news item, the unique attractor is $p_{A,k,x}$ is always much larger than $\frac{1}{2}$ as seen in Table (\ref{tt}) and  Fig. (\ref{px}). It means that even if fake news is first believed by only a handful of agents, the informal discussions among them, will inexorably increases the proportion of believers to end up with an overwhelming part of the community. This situation is most worrying with the fake news item reaching a stable status of being "true" within the related community.

\end{description}

\begin{figure}
\includegraphics[width=.50\textwidth]{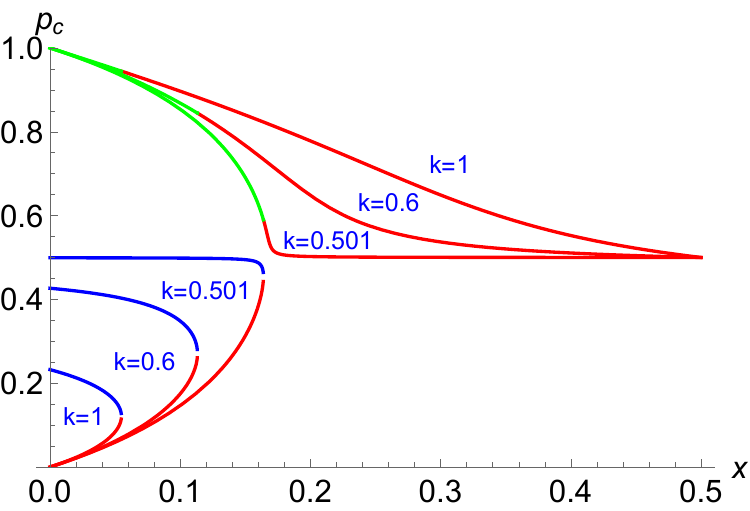}\quad
\includegraphics[width=.50\textwidth]{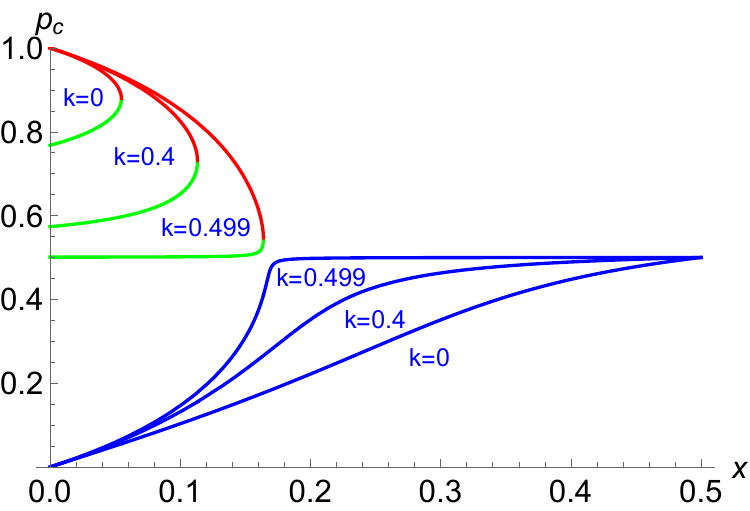}\\ \\ \\
\includegraphics[width=1\textwidth]{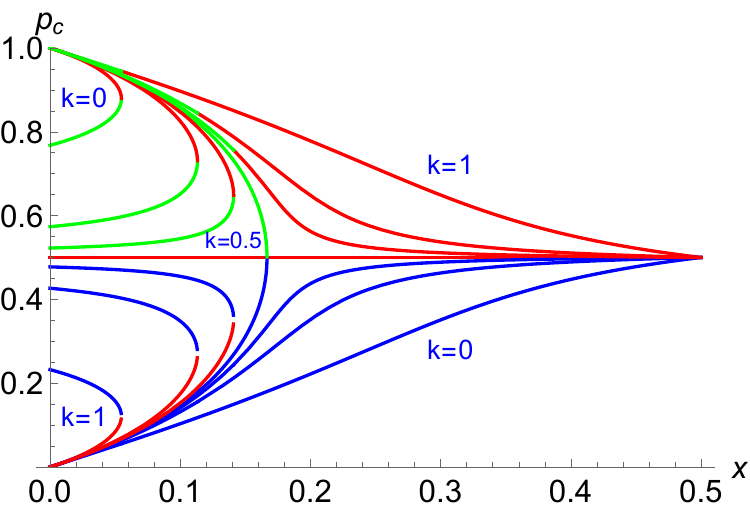}
\caption{Attractors and tipping poing  as a function of $x$ for $k=1, 0.6, 0.501$ and $k=0.499, 0.4, 0$ in the upper part respectively left and right parts. Lower part includes both series. The attractor $p_{B, k, x}$ is in blue), the attractor $p_{A, k, x}$ in red and the tipping point $p_{t, k, x}$ in green. All are denoted by the label $p_c$.}
\label{all}
\end{figure}

To grasp the whole landscape of the various types of dynamics I have aggregated the cases $k=1, 0.6, 0.501$ and $k=0.499, 0.4, 0$ in the upper part of Fig (\ref{all}), respectively left and right parts. Lower part includes both series.

\section{Strategies to sequestrate fake news without prohibiting them}

The series of results obtained highlight the critical roles played by contrarians and activated prejudices in shaping the fate of a given piece of fake news.  It is worth to stress that their respective roles are implemented in parallel and without interaction between them. The findings set the frame to identify new avenues for designing strategies to curb the prominent spread of fake news and sequestrate it naturally into small networks of agents without the need for prohibition. The key findings are:

\begin{description}
\item[(i)] 
A few percents of contrarians are enough to modify drastically the full landscape of the associated dynamics of spreading or shrinking as exhibited in Fig (\ref{all}). Contrarians have always the same impact, which is transforming the tipping point dynamic into that of a single attractor. The consequence is that any initial support for any fake news ends up at this unique attractor whose value is independent of that of the initial support.

\item[(ii)] The location of the unique attractor of the dynamics is either above or below $\frac{1}{2}$ depending on the distribution of prejudices activated by the fake news item. It is also worth to emphasize that the single attractor is located mostly at either very low or very high values for as a function of the distribution of activated prejudices.

\item[(iii)] The heterogeneities of the activated prejudices depend on the sociocultural composition of each community. Moreover, the prejudices, which are activated spontaneously, are selected by the content of the piece of fake news. As a result, some identical fake news may spread in some communities and shrink in others.

\end{description}

\subsection{Most fake news do not spread}

While contrarians are instrumental to the drastic reshaping of the geometry of the dynamics landscape, I notice that they need to reach a proportion ranging between 6 and 20 \% to turn the tipping point dynamics into a single attractor dynamics  (see Table (\ref{tt})). More than 10\%  is a significant figure, which in turn indicates that it is unlikely that many fake news would generate such proportions of contrarians. 

\begin{description}
\item[(i)] When the contrarians are few in number, even in the case of beneficial (prejudices $k=1$), fake news needs to start  with a rather high proportion of individual believers. also rather hard to reach as seen in Fig. (\ref{all}). For instance with $x=0$, the initial proportion must be higher than 23\%. However, there are very rare exceptions, such as the false claim that Israel bombed a hospital in Gaza in October 2023, which reached an impressive number of believers around the world in a matter of hours. \cite{fa1, fa2, fa3, fa4, fa5, fa6}.

\item[(ii)] When the fake news item is at odds with the prejudices, the challenge becomes out of reach. In the case $k=0$ and $x=0$, the initial support must be higher than 77\% (Fig. (\ref{all})). A huge figure impossible to reach in most cases. 

\item[(ii)] In cases where the fake news item does generate proportions of contrarians about 10\%, substantial proportions of initial believers are still necessary for both $k<\frac{1}{2}$ and $k>\frac{1}{2}$ as seen in Fig. (\ref{all}), 

\end{description}

All these observations indicate that most fake news do not spread and thus do not pose a threat to democratic unbiased public debates.

\subsection{Some rare fake news turn spontaneously invasive}

When the fake news item is able to simultaneously generate more than 15\% of contrarians and be in tune with most activated prejudices, only a handful of initial believers is sufficient to launch an invasive dynamics of opinion. Then, the proportion of supporters grows unseen till reaching eventually a majority of agents despite being initially dismissed as false by the majority of the community as shown in Fig. (\ref{all}),

It may also happen, although very rarely, that a fake news post is believed at once as true by almost everyone  \cite{fa1, fa2, fa3, fa4, fa5, fa6}. When that happens, if the fake news item is in tune with the activated prejudices, it stays widely believed through local discussions and settled as ``true" within the population, which in turn can lead to concrete actions by the agents \cite{fa1, fa2, fa3, fa4, fa5, fa6}. 

\subsection{Novel strategies to curb invasive fake news}
\label{risk}

Indeed, above findings and results indicate that to prevent malicious fake news thwarting the outcomes of balanced democratic public debates about central societal and political issues, the focus could be not on banning the making of fake news but on either prevent their spreading or drive their shrinking.

As long as a fake news item stays confined to a small group of agents it does not pose a threat. Therefore, instead of limiting the freedom of speech with heavy control of social media, it may be more efficient to address the geometry of the landscape of opinion, which drives the propagation of fake news. The subsequent goal becomes to identify the features, which keep most fake news confined and then apply them to the ones which do spread over. Accordingly, I advocate on setting the conditions for sequestrating invasive fake news within small networks of agents. In the rare cases where the initial proportion of believers is very high, the aim is to get the initial support reduced to small values and to keep it there.

Within the extended Galam model of opinion dynamics, two parameters can fulfill this goal of "No ban, No spread - with Sequestration". First one is $k$, the spectrum of prejudices activated spontaneously when groups of discussing agents get trapped at a tie with the two choices, true and fake, being equally acceptable. The second one is $x$, the proportion of contrarian agents which is produced by both the content of the fake news item and the social environment of the agents.

The values of these two parameters determine the geometry of the landscape in which the fake news dynamics is deployed. When the geometry boosted a fake news propagation, its reshaping requests to modify these values to get them set at very low values to activate the sequestration of fake news with no possibility to spread out. More specifically, the proportion of contrarians must be reduced, which implies individual changes of attitude. In addition, tie breaking cases must be tuned against the fake news item, which requires modifying the associated prejudices. In case of an initial high proportion of believers, turning $k$ from high to low values, triggers a shrinking dynamics.

Once this avenue is selected, the follow-up instrumental question is how to implement such a reshaping scheme within the real world. That requires elaborating appropriate novel tools to modify $k$ and $x$ in order to reshape the social geometry of the landscape of the dynamics of fake news.

However, defining the appropriate corresponding tools is beyond my skills and expertise. At this stage, I have shown that the scheme of "No ban, No spread - with Sequestration" is feasible within a stylized model of opinion dynamics. To transpose the related findings to the real world requires an interdisciplinary collaboration with scientists from Computer, Psychological, Cognitive and Behavioral sciences. I hope this paper will stimulate in these communities. an interest along this point of view.

\section{Conclusion}

In this paper I have extended the Galam model of opinion dynamics to study the dynamics of fakes news. In particular, I focused on the effects of combining prejudices tie breaking and contrarian behavior. These two parameters being independent of each other.

Solving the update equations has revealed the existence of a critical value for the proportion of contrarians, above which the dynamics landscape is turned up-side-down. There, the dynamics shifts abruptly from a tipping point dynamics to a single attractor dynamics. Remarkably, the related contrarian critical value is small with values between 6 to 20 \%. 

In parallel, the associated single attractor is found to be located at either high or very low proportions of fake news believers. The selection depends on the fake news item being in tune or at odds with the main activated prejudices. A high value means the fake news item does eventually invade the social space whatever its initial support, even with only a handful of agents. On the contrary, a low value means the fake ends up sequestrated into a very small proportion of believers even if it starts from an initial high support.

These findings are instrumental in determining the fate of fake news. Unveiling the social geometry of the opinion landscape has shown that most pieces of fake news never reach significant proportions of believers and stay confined to small numbers of agents. Therefore, they pose no threat.

However, the subgroup of fake news items, which generates enough contrarians and are in tune with most activated prejudices, do spread and invade large parts of a social community even when initially shared by only a handful of agents.  The phenomenon is counterintuitive and unexpected highlighting an alternative understanding of the ``natural" spread of fake news. 

Examining the fake news issue through this lens sheds a new light on the underlying processes, which makes a fake news invasive with no barrier. It thus allows envisioning new targeted paths to tackle only those threatening pieces of invasive fake news by sequestrating them via a modification of the underlying social geometry  to prevent their otherwise propagation towards a majority of believers. 

It is worth to emphasize that this approach is non restrictive. Neither ban nor control of the Net is required. If successful, fake news could prosper with no prohibition allowing total freedom of speech with simultaneously the guarantee of getting them naturally sequestrated into limited social networks involving only few numbers of agents.

This work illustrates the decisive role the interdisciplinary approach of sociophysics can provide to unveil a deeper understanding of the fake news phenomenon. The unexpected findings may contribute to the development of effective measures to address the issue of fake news, which otherwise could put at stake the democratic character of modern societies. 

At this stage, it is of importance to stress that allowing the existence of fake news per se does not mean allowing speech of hate, which are forbidden by law and thus must be prosecuted and condemned.

Last, but not least, I must make the following warning statement: in case, tools can be designed and implemented to monitor both the selection of activated prejudices and the production of contrarians, malicious uses will also become possible. To temper this risk, I would say that while the present situation is beneficial to malicious purposes of fake news, those new tools could be helpful to oppose and reverse the current balance of power, which is at the advantage of malicious fake news.

The risk of malicious use is inherent to scientific discoveries. The responsibility for their use and application lies with policy-makers and citizens alike. The responsibility of researchers is to ensure that their research is public and available.

\section*{Acknowledgement} I thank Kevin Arceneaux for his valuable suggestions and comments about the earlier draft of this article.

\end{document}